\documentclass[prd,preprint]{revtex4-2}
\usepackage{graphicx} 
\usepackage{amsmath,amssymb}
\usepackage{amsmath}
\usepackage{comment}
\usepackage{url}
\usepackage{xcolor}
\usepackage[normalem]{ulem}
\usepackage{orcidlink}

\begin{document}

\title{Quantizability of Lorentz-violating field theories with singular Hamiltonians}

\author{Md Saif \orcidlink{0009-0008-7578-9524}}
\email{saif\_2421ph04@iitp.ac.in}

\author{Sandeep Aashish \orcidlink{0000-0003-1316-5966}}
\email{aashish@iitp.ac.in}

\affiliation{Department of Physics, Indian Institute of Technology Patna, Patna, Bihar 801106, India}

\date{\today}

\begin{abstract}
Spontaneous Lorentz-violating models of an antisymmetric tensor field are known to possess a singular Hamiltonian on the vacuum manifold, leading to unresolvable pathologies that render such theories unfit for cosmological studies. In this work, we introduce an auxiliary vector field inspired by the St\"uckelberg mechanism to investigate whether this resolves the fundamental phase space singularities. The total Hamiltonian is found to be globally ill-defined. For completeness, in an effective field theoretic sense where the phase space is restricted to near the vacuum manifold of Lorentz-violating models, we derive the singularity conditions near the vacuum manifold for an explicit model and find that the resulting constraint matrix acquires an explicit dependence on the gradients and conjugate momenta of the St\"uckelberg field components that can still lead to a singular Hamiltonian.
\end{abstract}

\maketitle
\newpage

\section{Introduction}
A longstanding problem of interest in theoretical physics has been the possibility that small violations of Lorentz symmetry might reveal new fundamental physics that cannot be seen in current theories \cite{Colladay_96, Kostelecky_99, Kostelecky_88, Bardakci_77, Bardakci_74, Kostelecky_02}. In many models that preserve fundamental consistency conditions like the Bianchi identities \cite{Kostelecky_03}, such violations do not arise from an explicit breaking of Lorentz symmetry at the level of the action and equations of motion. Instead, the Lorentz violation is effected through a vector or a tensor field whose dynamics are constructed to remain fully Lorentz-invariant, but because of the way its dynamics are set up, the field naturally settles into its lowest-energy state with a non-zero value. Since this field carries directional information, having a non-zero value means that the ground state itself now has a preferred frame of reference, and as a result, Lorentz symmetry is broken spontaneously \cite{Kostelecky_88, Seifert_09, Bluhm_07}. 

One of the first spontaneous Lorentz-violating scenarios using vector fields constitutes the so-called bumblebee models, in which a vector field acquires a non-zero vacuum expectation value (VEV) through a Higgs-like potential \cite{Kostelecky_88, Bluhm_07}. These vector-field models have been studied in both flat spacetime as well as in the presence of gravity \cite{Kostelecky_03, Hernask_16, Kostelecky_09, Bailey_06, Seifert_07, Liu_25}. 
Beyond vector fields, higher-rank tensor fields have also been explored as sources of Lorentz symmetry breaking. Tensor fields allow for more general symmetry-breaking patterns, as they can select preferred planes or more complicated structures in spacetime rather than a single direction. These studies have led to a general effective field-theoretic framework for spontaneous Lorentz violation known as the Standard-Model Extension (SME) \cite{Kostelecky_99, Colladay_96, Kostelecky_03, Kostelecky_05}, which parametrizes all possible Lorentz- and CPT-violating operators consistent with observer covariance. Within the SME, the coefficients for Lorentz violation are interpreted as vacuum expectation values of underlying tensor fields, thereby providing a unified description that connects symmetry breaking, field dynamics, and phenomenology across particle physics and gravity.

Among the higher rank tensor field models, the rank-two antisymmetric tensor field $B_{ab}$ (or Kalb-Ramond field) is exciting because it is well motivated from a theoretical point of view, emerging naturally in the low-energy limit of superstring models \cite{Kalb_74, Fradkin_84, Callan_85, Gasperini_92}. In low-energy effective descriptions, these fields come with a standard gauge-invariant kinetic term built from a completely antisymmetric field strength \cite{Kalb_74, Aashish_18, Aashish_19, VanNieuwenhuizen_73, Aashish_19A, Ajith_22}. Models based on such antisymmetric tensor fields have been studied to understand spontaneous Lorentz violation induced by Higgs-like potentials \cite{Altschul_09, Assuncao_19, Aashish_19, Aashish_21, Ajith_22}. In these models, the potential is chosen in such a way that it involves only those combinations of the tensor field that are Lorentz invariant so that the field acquires a preferred vacuum configuration, given by $\langle B_{ab}\rangle = b_{ab}$, without breaking Lorentz symmetry of the equation of motion.

More recently, attention has shifted toward the canonical structure of these theories. It has been shown that both vector and antisymmetric tensor field models with spontaneous Lorentz violation can exhibit nontrivial problems at the Hamiltonian level \cite{Seifert_19, Seifert_18}. In particular, Seifert \cite{Seifert_19} showed that in models where Lorentz symmetry is broken spontaneously by vector or tensor fields with Higgs-like potentials, the Hamiltonian can become ill-defined on the set of vacuum solutions. When this happens, the equations that govern time evolution fail to remain consistent in the ground state of the theory, and the theory cannot consistently describe small fluctuations around it. Moreover, the constraint and invariant structure of antisymmetric tensor field models are such that this pathology cannot be resolved by constructing special kinetic or potential terms of $B_{ab}$, unlike in vector field models \cite{Seifert_19}.

These studies provide an important starting point for studying similar questions in theories involving higher-rank tensor fields. In the past, the Hamiltonian analysis of Lorentz-violating bumblebee models has been extensively studied in Refs. \cite{Seifert_09, Seifert_19}. Hernaski \cite{Hernask_16} studied spontaneous Lorentz symmetry breaking with rank-two antisymmetric tensor and analyzed the corresponding low-energy theory, including the properties of its Goldstone modes, stability, and causality. More recently, Potting \cite{Potting_23} examined the Hamiltonian stability of models containing an antisymmetric two-tensor. It was shown that a quadratic model involving only the antisymmetric tensor does not lead to a Hamiltonian bounded from below, but a so-called hybrid model, in which an additional vector field is coupled to the antisymmetric tensor field, can yield a ghost-free Hamiltonian in Minkowski spacetime. These results show that the presence of an additional vector sector can play an important role in the Hamiltonian structure of antisymmetric tensor models. However, the effect of such an extension on the constraint structure and on the possible singularity of the canonical formulation remains unaddressed. Moreover, the motivation, or rather a physically motivated prescription for constructing such ``hybrid" models, is not yet addressed in the literature.
 
Analogous to the past works in Ref.~\cite{Hernask_16, Potting_23}, this paper systematically explores how adding this vector sector via the St\"uckelberg mechanism changes the constraint structure and affects the singularities of the canonical formulation.
One possibility is to explore how the issue of singular Hamiltonians at the classical level generalizes in the context of the quantization of Lorentz-violating models, also noted as one of the limitations of Ref. \cite{Seifert_19}. The quantization of such models involves the use of the St\"uckelberg mechanism \cite{Stueckelberg_38-1, Stueckelberg_38-2, RueggRuizAltaba_04, Buchbinder_08, Hinterbichler_12, Govindarajan_25} wherein a so-called St\"uckelberg field (a vector field $C_{a}$ in case of antisymmetric tensor field model) is introduced in the Lagrangian to restore the gauge symmetry so as to apply standard quantization procedures like the Faddeev-Popov method \cite{Faddeev_67, Peskin_95}. It is therefore interesting to study how the singularity of Hamiltonian is affected once a pathological model of antisymmetric tensor field is treated with the St\"uckelberg mechanism.

In this work, we consider a class of St\"uckelberg mechanism-inspired models of antisymmetric tensor field with spontaneously Lorentz-violating potential limited to quadratic order derivatives to investigate this pathology and therefore quantizability of such models. For completeness, we also consider, in an effective field theoretic sense \cite{Altschul_09, Kostelecky_88, Kostelecky_03, BluhmRobert_08}, a restricted phase space corresponding to small fluctuations around the vacuum manifold and comment on the conditions for singularity of the constraint matrix.

The paper is organized as follows. Sec. \ref{sec2} provides a brief review of the elements of Hamiltonian field theory and St\"uckelberg mechanism that are needed for the analysis. In Sec.~\ref{sec3}, we study a Lorentz-violating field theory based on a rank-two antisymmetric tensor field $B_{ab}$ with a general potential term treated with the St\"uckelberg procedure to restore gauge invariance of the action. The general structure of the coefficient matrix of Lagrange multipliers is obtained. In Sec.~\ref{sec4}, we consider an explicit model with up to quadratic order derivative terms and evaluate the inverse of the coefficient matrix $M_{ij}$ for a specific choice of the vacuum value of $B_{ab}$. Finally, in Sec. \ref{sec5} we conclude with a few remarks about the results.

\section{Dirac-Bergmann analysis and the St\"uckelberg mechanism}
\label{sec2}

In this section, we will briefly review the Dirac-Bergmann analysis procedure to study the singularities in the Hamiltonian. For a comprehensive review, the interested reader(s) may go through the Refs. \cite{Dirac_64, Isenberg_77, Brown_22, Garfinkle_12, Lueiza_26}. Throughout this work, we follow the notations and conventions introduced in Ref. \cite{Seifert_19}. The coordinate dependence of fields and functions is implicit in all subsequent expressions, unless stated otherwise.

We consider a Lagrangian density
$\mathcal{L}(\psi_{\alpha}, \dot{\psi}_{\alpha}, \nabla_i \psi_{\alpha})$. In regular systems, the Legendre transformation, i.e., the Hamiltonian density $\mathcal{H}_0 = \pi_\alpha \dot{\psi}_\alpha - \mathcal{L}$ is invertible, allowing all velocities to be expressed in terms of the phase-space variables $(\psi_\alpha,\pi_\alpha)$. However, in singular systems, the Hessian matrix is degenerate, and the inversion is not possible. As a consequence, one obtains relations among $(\psi_\alpha,\pi_\alpha)$ that do not involve time derivatives. These relations restrict the allowed phase-space configurations and are called \textit{primary constraints} and can be written in the general form as
$\Phi_I(\psi_\alpha, \nabla \psi_\alpha, \pi_\alpha) = 0$ where, $\quad I=1,\dots,N$.

The Dirac--Bergmann algorithm proceeds as follows:

\begin{enumerate}

\item[(i)]  
To ensure that the constraints hold at all times, one defines the augmented Hamiltonian density as $\mathcal{H}_A = \mathcal{H}_0 + u_I\,\Phi_I$, where $u_I (\mathbf{x},t)$ are arbitrary Lagrange multipliers. Their arbitrariness reflects the fact that the time evolution is not uniquely determined at this stage.

\item[(ii)]  
Time evolution of any phase-space function $f$ is generated by Poisson brackets with the total Hamiltonian $H_A\equiv \int d^3x \mathcal{H}_A$,
\begin{equation}
\dot{f} (\mathbf{x}) = \{f ( \mathbf{x}),H_A\} = \int d^3y \{f (\mathbf{x}),\mathcal{H}_A (\mathbf{y}) \}.
\end{equation}
For field theories, the Poisson bracket is defined as,
\begin{equation}
\{f (\mathbf{x}),H_A\} = \sum_\alpha \int d^3y \int d^3z 
\left[
\frac{\delta f (\mathbf{x}) }{\delta \psi_\alpha(\mathbf{z})}
\frac{\delta \mathcal{H}_A \mathbf{(y)}}{\delta \pi_\alpha(\mathbf{z})}
-
\frac{\delta f (\mathbf{x})}{\delta \pi_\alpha(\mathbf{z})}
\frac{\delta \mathcal{H}_A \mathbf{(y)}}{\delta \psi_\alpha(\mathbf{z})}
\right],
\end{equation}
where $\alpha$ runs over all fields and their corresponding conjugate momenta.
\item[(iii)]
Consistency requires that all primary constraints be preserved in time,
\begin{equation}
\dot{\Phi}_I(\mathbf{x}) = \{\Phi_I(\mathbf{x}), H_A\} \approx 0,
\end{equation}
where $\approx$ denotes equality on the constraint surface. If these conditions do not vanish identically, they lead to new relations among the phase-space variables, known as \textit{secondary constraints},
\begin{equation}
\Psi_I(\psi_\alpha, \nabla \psi_\alpha, \pi_\alpha) = 0.
\end{equation}

The consistency procedure must then be repeated by imposing
$\{\Psi_I(\mathbf{x}), H_A\} \approx 0$.
This iterative process may generate further (tertiary, etc.) constraints or determine some or all of the Lagrange multipliers $u_I$.
\end{enumerate}

The final Hamiltonian $H_A$ generates a consistent time evolution on the constraint surface. The number of physical degrees of freedom is obtained by removing both second-class constraints and gauge redundancies associated with first-class constraints. The undetermined Lagrange multipliers correspond to gauge freedom and do not represent physical degrees of freedom. Crucially, the transition from this classical constrained system to a consistent quantum theory relies on replacing Poisson brackets with Dirac brackets. The formulation of Dirac brackets dictates that the matrix composed of the Poisson brackets of all second-class constraints must be globally invertible. If this constraint matrix exhibits dynamical degeneracies or is globally singular, the fundamental global symplectic structure is lost, thereby obstructing standard canonical quantization.

\subsection{St\"uckelberg mechanism and the Lorentz-violating model}
The St\"uckelberg mechanism provides a systematic way to restore gauge invariance in theories where it is broken by mass terms or interactions depending explicitly on the gauge potential. By enlarging the field content with an auxiliary scalar field that compensates the gauge variation, the theory can be written in a gauge-invariant form without introducing additional physical degrees of freedom \cite{RueggRuizAltaba_04, Govindarajan_25, GrosseKnetter_93}. This formulation should be understood as a gauge-covariant rewriting of the original theory.

We consider an antisymmetric tensor field $B_{ab} = - B_{ba}$ model given by the Lagrangian density,
\begin{equation}
\label{eq_genL}
\mathcal{L}
= -\frac{1}{12} H_{abc} H^{abc}
- \lambda V(B_{ab}) ,
\end{equation}
where $H_{abc} = 3\,\partial_{[a} B_{bc]} $ and $\lambda$ is the coupling strength. The potential term $V(B_{ab})$ is constructed from the invariants $X=B_{ab}B^{ab}$ (parity-even) and $Y=\frac{1}{2} \epsilon^{abcd}B_{cd} B_{ab} $ (parity-odd) \cite{Seifert_19}. For our purposes, we restrict to parity-even terms in the potential. Also relevant to our study is the fact that in order to support Lorentz symmetry breaking, the antisymmetric tensor field $B_{ab}$ must admit vacuum solutions in which it acquires a constant, non-zero value \cite{Seifert_19} and the simplest parity-even potential term consists of quartic order terms of the field $B_{ab}$ or, equivalently, quadratic order terms in $X$. Note that the kinetic term in Eq. (\ref{eq_genL}) is symmetric under the transformation:
\begin{equation}
\label{eq:gauge_B}
B_{ab}
\longrightarrow
B'_{ab}
=
B_{ab}
+
\partial_a \Lambda_b -
\partial_b \Lambda_a ,
\end{equation}
where $\Lambda_a$ is an arbitrary vector gauge parameter. However, unlike an ordinary vector gauge field, the gauge parameter $\Lambda_a$ itself possesses additional gauge-like symmetry given by,
\begin{equation}
\label{gparamtrans}
\Lambda_a\to \Lambda_a^{'}= \Lambda_a + \partial_a\omega,
\end{equation}
where $\omega$ is an arbitrary scalar function, that leaves the field $B_{ab}$ invariant. In this case, the standard quantization approaches for gauge theories like the Faddeev-Popov method are known to fail to consistently quantize the theory \cite{Buchbinder_07, Barvinsky_85} and are generally categorized as a class of reducible gauge theories.
Batalin and Fradkin \cite{Batalin_83} developed a generalized canonical formalism for reducible gauge theories. A more general framework for the quantization of gauge theories with linearly dependent gauge generators was developed by Batalin and Vilkovisky \cite{BatalinVilkovisky_83}. More recently, a geometrical prescription for dealing with such theories was developed in Ref. \cite{Aashish_18b}.

Additionally, the Lagrangian (\ref{eq_genL}) consists of a potential term that is not invariant under the transformation Eq. (\ref{eq:gauge_B}). Therefore, the Lagrangian \eqref{eq_genL} belongs to a class of models with a gauge-invariant kinetic term but a gauge-breaking potential term. It turns out that, despite the Lagrangian \eqref{eq_genL} itself being not gauge invariant, its quantization through the methods described above leads to divergences attributed to the gauge invariance of the kinetic term \cite{Buchbinder_08, RueggRuizAltaba_04}. This issue of gauge-related divergences can be resolved by restoring the gauge symmetry through the introduction of a so-called St\"uckelberg field. For example, in the case of an antisymmetric tensor field model with the above characteristics, one introduces a St\"uckelberg vector field $C_a$ with field strength
\begin{equation}
F_{ab} = \partial_a C_b - \partial_b C_a ,
\end{equation}
and replaces $B_{ab} \rightarrow B_{ab} + \frac{1}{m} F_{ab}$ to restore the gauge symmetry in the potential term, where $m$ is the parameter introduced to balance the derivative appearing in $F_{ab}$. For instance, in the case where $V(B_{ab})=(B_{ab}B^{ab}-b^2)^2$, the St\"uckelberg-extended Lagrangian,
\begin{equation}
\label{eq:lagrangian_st}
\mathcal{L}_{st}
=
-\frac{1}{12} H_{abc} H^{abc}
-\lambda\!\left[(B_{ab}+\frac{1}{m} F_{ab})(B^{ab}+ \frac{1}{m} F^{ab})-b^2 \right]^2 ,
\end{equation}
is invariant under the transformations
\begin{align}
\label{gaugetr}
B_{ab} &\rightarrow B_{ab}
+\partial_a \Lambda_b - \partial_b \Lambda_a \ ;
&
C_a &\rightarrow C_a - m \Lambda_a. 
\end{align}

In the next section, we perform the Hamiltonian constraint analysis for a spontaneously Lorentz-violating model of $B_{ab}$ with a general parity-even potential term with up to quadratic order derivatives.

\section{Dirac-Bergmann analysis of Lorentz-violating antisymmetric tensor field model}
\label{sec3}

We start by considering the Lagrangian $\mathcal{L}_{st}$ given by Eq. (\ref{eq:lagrangian_st}), describing a St\"uckelberg treated model of a rank-two antisymmetric tensor field with spontaneous Lorentz violation in a Minkowski background. To construct the Hamiltonian of this model, we perform a $3+1$ decomposition, wherein the antisymmetric tensor field $B_{ab}$ can be decomposed into two spatial vector fields $\vec{P}$ and $\vec{Q}$, which play roles analogous to electric and magnetic components of the electromagnetic field strength tensor. These are defined as \cite{Seifert_19},
\begin{equation}
P^{i} \equiv B^{0i}, \qquad 
Q^{i} \equiv \frac{1}{2}\,\epsilon^{ijk} B^{jk} , 
\end{equation}
where $\epsilon_{ijk}$ denotes the Levi--Civita symbol on a constant-$t$ spatial hypersurface in spacetime.

Before we proceed, a comment is in order. A general analysis in the present context would, in principle, consider a curved spacetime background instead of Minkowski spacetime. The Kalb--Ramond field has indeed been studied in curved spacetime in literature, both in the context of spontaneous Lorentz violation and in its gravitational and cosmological applications \cite{Maluf_22, Aashish_19, Lessa_19}. In particular, quantum aspects of an antisymmetric tensor field with spontaneous Lorentz violation have been studied in curved spacetime using a covariant effective-action approach~\cite{Aashish_19, Buchbinder_08}. Several studies have subsequently explored the gravitational effects of Kalb--Ramond backgrounds, including modified black-hole solutions \cite{Lessa_19}, particle dynamics and gravitational lensing around Kalb--Ramond black holes~\cite{Atamurotov_22, JuniorEdnaldo_24}, and cosmological solutions in which the Kalb--Ramond background can lead to direction-dependent expansion~\cite{Maluf_22}. In the present work, however, we limit our analysis to flat spacetime, while the corresponding Hamiltonian analysis in a curved background is left for future work.

In terms of the fields $P^{i}, Q^{i}$, the kinetic term appearing in the Lagrangian can be rewritten as,
\begin{align}
    -\frac{1}{12} H_{abc} H^{abc}
= \frac{1}{2} \Big[
( \dot{\vec{Q}} - \vec{\nabla} \times \vec{P} )^{2}
- ( \vec{\nabla} \cdot {\vec{Q}} )^{2}
\Big].
\end{align}

The potential term in Eq. (\ref{eq:lagrangian_st}) generates a number of interaction terms involving the kinetic term of the St\"uckelberg field $F_{ab}$ coupled to $B_{ab}$:
\begin{align}
\label{eq:pot}
V_{st} = &
(B_{ab}B^{ab})^{2} + 
\frac{1}{m^{2}}\Big(B_{ab}F^{ab}\Big)^{2} + 
\frac{1}{m^{2}}\Big(F_{ab}B^{ab}\Big)^{2}+ 
\frac{1}{m^{4}}\Big(F_{ab}F^{ab}\Big)^{2}   
+ b^{4} \nonumber \\
&
+ \frac{2}{m} B_{ab} B^{ab} B_{cd} F^{cd}
+ \frac{2}{m} B_{ab} B^{ab} F_{cd} B^{cd} + 
\frac{2}{m^{2}} B_{ab} B^{ab} F_{cd} F^{cd} \nonumber \\
& - 2 b^2 B_{ab} B^{ab} 
+ \frac{2}{m^{2}} B_{ab} F^{ab} F_{cd} B^{cd} +
\frac{2}{m^{3}} B_{ab} F^{ab} F_{cd} F^{cd} - 
\frac{2 b^2}{m} B_{ab} F^{ab}  \nonumber \\
& + \frac{2}{m^{3}} F_{ab} B^{ab} F_{cd} F^{cd} - 
\frac{2 b^2}{m} F_{ab} B^{ab} - 
\frac{2 b^2}{m^{2}} F_{ab} F^{ab} .
\end{align}

Notice the presence of quartic and third-order terms of $F_{ab}$ in Eq. (\ref{eq:pot}), resulting in higher than quadratic order derivatives in the Lagrangian. In principle, such higher-order derivative terms lead to instabilities like the Ostrogradsky instability that can be avoided, at least in certain conditions, using existing approaches incorporating quantization \cite{Woodard_15, Donoghue_21, Pais_50}. However, to avoid such complications, we restrict the potential term in Eq. (\ref{eq:pot}) to terms up to quadratic order in derivatives by neglecting $\mathcal{O}(1/m^3)$ terms. 
This choice is justified in the sense that terms proportional to $1/m^{3}$ and $1/m^{4}$ correspond to higher-order interactions suppressed by higher powers of the mass scale ($m^{-3}, m^{-4}$) and can be neglected within an effective field theory (EFT) treatment in the limit where the magnitude $\|m\| \gg \|F_{ab}\|$. 

Since the St\"uckelberg field $C_a$ only comes into play in the theory through its field strength $F_{ab}$, any contribution to the Lagrangian that is quadratic in $C_a$ also involves derivatives of this field. For example, $F_{0i} = \dot C_i - \partial_i C_0$. These terms therefore contribute to the kinetic term of the model in Eq. (\ref{eq:lagrangian_st}). Hence, the model under consideration in this work is given by,
\begin{align}
\label{Symbolic-Lagrangian}
\mathcal{L}
&= -\frac{1}{12} H_{abc} H^{abc}
   - \lambda\left[
   \frac{1}{m^2}\,k_1(\vec P,\vec Q; b^2)\,\dot{\vec C}^{\,2}
   + \frac{1}{m^{2}}\,k_2\,(\vec P\cdot\dot{\vec C})^{2} + 
   V\!\left(B_{ab}, F_{ab}; b^2 \right)
   \right].
\end{align}
Here, the coefficients $k_1(\vec P,\vec Q; b^2)$ and $k_2$ are the generalized coefficients of the kinetic terms $\dot{\vec C}^{\,2}$ and $(\vec P\cdot\dot{\vec C})^{2}$ respectively, and may be functions of fields and their gradients for a general potential $V_{st}(B_{ab})$. 

\subsection{Hamiltonian analysis}

Now, from the Lagrangian \eqref{Symbolic-Lagrangian}, we can find the conjugate momenta corresponding to the fields $\vec{P}$, $\vec{Q}$, $C_{0}$ and $\vec{C}$ as,
\begin{eqnarray}
    \vec{\Pi}_{P} &=& \frac{\delta \mathcal{L}}{\delta \dot{\vec{P}}}=0 ,\nonumber \\
    \vec{\Pi}_{Q} &=& \frac{\delta \mathcal{L}}{\delta \dot{\vec{Q}}}= \dot{\vec{Q}} - \vec{\nabla} \times \vec{P} , \nonumber \\
    \dot{\vec{Q}} &=& \vec{\Pi}_{Q} + \vec{\nabla} \times \vec{P} ,\\
    \Pi_{C_{0}} &=& \frac{\delta \mathcal{L}}{\delta \dot{C_{0}}} = 0 , \nonumber \\
    \vec{\Pi}_{C} &=& \frac{\delta \mathcal{L}}{\delta \dot{\vec{C}}} = 
    - \lambda \Big[\frac{2k_{1}}{m^2}\dot{\vec{C}} + 
    \frac{2k_{2}}{m^{2}} (\vec{P} \cdot \dot{\vec{C}})\vec{P} + 
    \lambda \frac{\partial V}{\partial \dot{\vec{C}}}
    \Big] \nonumber.
\end{eqnarray}
For $\vec{\Pi}_C$, after some algebraic manipulations we find,
\begin{equation}
\label{cdot}
\dot{\vec C}
= -\,\frac{\vec\Pi_C + \lambda\,\vec{V}_{,\dot{\vec C}}}{k_0} , 
\end{equation}
where, we have taken $\vec{V}_{,\dot{\vec C}} \equiv \frac{\partial V}{\partial \dot{\vec C}}$ and $k_0 \;\equiv\; \frac{2\lambda}{m^2} \!\left(k_1+k_2 \vec{P}^{2}\right)$. Note that, in a general phase space, the denominator $k_0$ in Eq. (\ref{cdot}) may vanish and therefore the augmented Hamiltonian is globally ill-defined, preventing consistent quantization of the theory. However, in an effective field theoretic sense, spontaneous Lorentz-violating models are studied in the vicinity of the vacuum value (or the vacuum manifold in phase space) \cite{Altschul_09, Buchbinder_08, BluhmRobert_08}, so it is still relevant to perform the Hamiltonian analysis in a restricted phase space where the EFT in question is valid, i.e., in phase space near the vacuum manifold. In what follows, we continue the constraint analysis strictly in this effective field theoretic sense. 

The constraint structure consists of four primary constraints. Three of them arise from the vanishing of the momenta conjugate to the spatial components of the vector field $\vec{P}$, which can be collected into the vector constraint
\begin{equation}
\vec{\Phi} \equiv \vec{\Pi}_{P} = 0 . 
\end{equation}

In addition, there is a scalar primary constraint associated with the momentum conjugate to the field $C_{0}$, given by
\begin{equation}
\Phi_{C_{0}} \equiv \Pi_{C_{0}} = 0 . 
\end{equation}

To ensure that all these constraints are preserved under time evolution, the Hamiltonian must be supplemented by appropriate Lagrange multipliers. The three multipliers enforcing $\vec{\Phi}$ are conveniently grouped into a vector $\vec{u}$, while the scalar constraint $\Phi_{C_{0}}$ is enforced by an independent multiplier $w$. With this choice of notation, the augmented Hamiltonian can be written in a clear and compact form, making the role of each constraint explicit as,
\begin{align*}
    \mathcal{H}_{A}= \vec{\Pi}_{P} \cdot \dot{\vec{P}} +  \vec{\Pi}_{Q} \cdot \dot{\vec{Q}} + \Pi_{C_{0}}\dot{C_{0}} +  \vec{\Pi}_{C} \cdot \dot{\vec{C}} - \mathcal{L} + \vec{u} \cdot  \vec{\Pi}_{P} + w \Pi_{C_{0}} . 
\end{align*}
As $\vec{\Pi}_{P}=0$ and $\Pi_{C_{0}}=0$ then $\mathcal{H}_{A}$ becomes
\begin{align}
\label{Hamiltonian_1}
    \mathcal{H}_{A} 
    \;=\;& \frac{1}{2} \vec{\Pi}_{Q}^2 - 
    \frac{\vec{\Pi}_{C}^2}{k_0} +
     \frac{\lambda k_1}{m^2 k_0^2} \vec{\Pi}_{C}^2+
     \frac{\lambda k_2}{m^2 k_0^2} (\vec{P} \cdot\vec{\Pi}_{C})^2+
    \vec{\Pi}_{Q} \cdot (\vec{\nabla} \times \vec{P})+ 
    \nonumber \\
    & \frac{1}{2} (\vec{\nabla} \cdot \vec{Q})^{2} + 
    V_{\text{eff}} +
    \vec{u} \cdot \vec{\Pi}_{P} + w\Pi_{C_{0}} , 
\end{align}
where $V_{\text{eff}}$ is
\begin{align}
    V_{\text{eff}}  \;=\;&
    \frac{\lambda^3 k_1}{m^2 k_0^2}  \vec{V}_{,\dot{\vec{C}}}^2 + 
    \frac{\lambda^3 k_2}{m^2 k_0^2} (\vec{P} \cdot \vec{V}_{,\dot{\vec{C}}})^2 +  \left(\frac{2\lambda^2 k_1}{m^2 k_0^2} - \frac{\lambda}{k_0}\right) (\vec{\Pi}_{C} \cdot \vec{V}_{,\dot{\vec{C}}}) + \nonumber \\
    & \frac{2\lambda^2 k_2}{m^2 k_0^2} (\vec{P} \cdot \vec{\Pi}_{C}) (\vec{P} \cdot \vec{V}_{,\dot{\vec{C}}}) + \lambda V .
\end{align}

\subsection{Constraint matrix}

We now require that the primary constraints be preserved under the time evolution generated by the Hamiltonian. This is implemented by imposing the consistency conditions
\begin{equation}
\dot{\Phi}_{a}=\{\Phi_{a}, H\} = 0 ,
\end{equation}
where $\{,\}$ denotes the Poisson bracket, and there is one such condition for each value of the index $a$ with $a = x, y, z, C_0$. These conditions ensure that the primary constraints $\Phi_{a}=0$ remain valid at all times. When the consistency conditions are evaluated on the surface defined by the primary constraints, three different possibilities can arise.
$(i)$ they may reduce to identities that are automatically satisfied once the primary constraints are enforced;
$(ii)$ they may impose conditions on the Lagrange multipliers, thereby fixing some of them in terms of the canonical variables $q$, $p$, and any multipliers that remain undetermined;
$(iii)$ they may lead to additional restrictions directly on the phase-space variables, resulting in new constraints on the system. The latter case corresponds to the emergence of genuinely new constraints, known as secondary constraints, which have the form $\Psi_a=0$.   
\begin{align}
    0=\dot{\vec{\Phi}} 
    \;=\;& 
    \{\vec{\Pi}_{P}, H_{A} \}  
    \nonumber \\
    \;=\;& 
    \frac{\partial}{\partial \vec{P} }
    \Big \{\frac{1}{k_0}\vec{\Pi}_{C}^2  -
    \frac{\lambda k_1}{m^2 k_{0}^2} \vec{\Pi}_{C}^2  - 
    \frac{\lambda k_2}{m^2 k_{0}^2} (\vec{P}  \cdot \vec{\Pi}_{C} )^2 
    \nonumber \\ 
    &- \vec{\Pi}_{Q}  \cdot (\vec{\nabla} \times \vec{P} ) -
    V_{\text{eff}} 
    \Big \} \equiv \vec{\Psi}  ,
    \label{SecondaryConstraint} \\
   0=\dot{\Phi}_{C_{0}} = & \{\Pi_{C_{0}} , H_{A} \} 
   \nonumber \\
    = & -\frac{\partial V_{\text{eff}}  } {\partial C_{0} }  \equiv \Psi_{C_{0}} .
   \label{SC_SBF}
\end{align}

The primary constraints $\vec{\Phi}$ and $\Phi_{C_{0}}$ are not preserved under time evolution, leading to secondary constraints $\vec{\Psi}$ and $\Psi_{C_{0}}$. Additionally, the secondary constraints must also be preserved as the system evolves over time. Imposing this condition leads to an equation that involves the previously undetermined Lagrange multipliers $\vec{u}$ and $w$. 

When the time evolution of the secondary constraint $\dot{\Psi}_{C_{0}}$ is evaluated, it is found that the Poisson bracket of the secondary constraint $\dot{\Psi}_{C_{0}}$ vanishes for the choice of potential Eq. (\ref{eq:potquad}) considered in this work, indicating that no further constraints arise in the theory. We demonstrate this by evaluating one of the Poisson brackets with respect to $(C_0,\Pi_{C_0})$, i.e., $\{\Psi_{C_{0}}, H_{A}\}$, which in principle would determine the Lagrange multiplier $w$: at quadratic order, the dependence of the effective potential on the temporal St\"uckelberg component $C_{0}$ involves only its spatial derivatives $\partial_i C_{0}$. We therefore rewrite the effective potential $V_{\text{eff}}$ in terms of orders of $C_{0}$,
\begin{equation}
\label{veff}
V_{\text{eff}}
=
\chi^1 (\vec{P},\vec{Q})\,
\partial_j C_{0} \partial_j C_{0}
+
\chi_{i}^2 \!\left(\vec{P},\vec{Q}, \,\vec{\nabla} \times \vec{C}, \vec{\Pi}_C \right)
\,\partial_i C_{0} + \chi^3(\vec{P}, \vec{Q}, \vec{\nabla} \times \vec{C}, \vec{\Pi}_C).
\end{equation}

Here $\chi^1, \chi^2 $ represent generalized coefficients with functional dependence on fields as shown above, and $\chi^3$ consists of terms independent of $C_0$.

Substituting Eq. (\ref{veff}) in Eq. (\ref{SC_SBF}), we obtain,
\begin{equation}
    \Psi_{C_{0}}= -\Big[\partial_i \Big(\chi^1 (\vec{P},\vec{Q})\,
(\partial_i C_{0}) \Big ) 
+
\partial_i \Big( \chi_{i}^2 \!\left(\vec{P},\vec{Q}, \,\vec{\nabla} \times \vec{C}, \vec{\Pi}_C \right) \Big ) \Big].
\end{equation}

With some algebraic manipulations, it can be shown that,
\begin{eqnarray}
    \{\Psi_{C_{0}},H_{A}\}_{C_{0}}
    =   \partial_{i}^2 \Big( \chi^1 (\vec{P},\vec{Q})\Big)
 -  \partial_{i}^2 \Big( \chi^1 (\vec{P},\vec{Q}) \Big) = 0.
\end{eqnarray}

Hence, the Lagrange multiplier $w$ is undetermined and can in turn be associated with the gauge symmetry of the theory under the transformations Eq. (\ref{gaugetr}). For a comprehensive quantization procedure, one must impose an appropriate gauge-fixing condition to uniquely determine this multiplier and convert first-class constraints into second-class ones. However, the fundamental pathology investigated in this work resides in the invertibility of the matrix governing the spatial components, $M_{ij}$, which we derive below. 

For the remaining secondary constraints, we get,
\begin{eqnarray}
     0=\dot{\Psi}_{i}=  \Big\{\Psi_{i},H_{A}\Big\} .
\end{eqnarray}
\begin{align}
    0=\dot{\Psi}_i =&
     \int d^{3}y\, 
\Big[
     \frac{\partial}{\partial P_{j} (\mathbf{y})}
     \Big(
   \frac{\partial}{\partial P_{i} (\mathbf{x})}
    \Big \{\frac{1}{k_0}\vec{\Pi}_{C}^2 (\mathbf{x}) -
    \frac{\lambda k_1}{m^2 k_{0}^2} \vec{\Pi}_{C}^2 (\mathbf{x}) - 
    \frac{\lambda k_2}{m^2 k_{0}^2} (\vec{P} (\mathbf{x}) \cdot \vec{\Pi}_{C} (\mathbf{x}) )^2 \nonumber \\ &
    - V_{\text{eff}} (\mathbf{x}) 
    \Big \}
    \Big)
    \Big] u_j (\mathbf{y}) + v_i (\mathbf{x}).
\end{align}

Since only the coefficients of the Lagrange multipliers $u_j$ are relevant for our further analysis, we therefore define the vector $v_i$ as the part of $\dot{\Psi}_i$ that is independent of the Lagrange multipliers,
\begin{equation}
v_i \equiv \int d^3y \ \{\Psi_i (\mathbf{x}) , \mathcal{H}_A\}.
\end{equation}
In symbolic form, this Poisson bracket can be written as,
\begin{equation}
v_i
=
\sum_{k} \int d^{3}y \int d^{3}z\,
\left[
\frac{\delta \Psi_i (\mathbf{x})}{\delta \zeta_{k}(\mathbf{z})}
\frac{\delta \mathcal{H}_{A}(\mathbf{y})}{\delta \Pi_{k}(\mathbf{z})}
-
\frac{\delta \Psi_i (\mathbf{x})}{\delta \Pi_{k}(\mathbf{z})}
\frac{\delta \mathcal{H}_{A} (\mathbf{y})}{\delta \zeta_{k} (\mathbf{z})}
\right],
\end{equation}
where the summation runs over the canonical field variables
$\zeta_k= \{ \vec{Q},C_0,\vec{C} \}$ and their conjugate momenta
$\Pi_k=\{ \vec{\Pi}_Q,\Pi_{C_0},\vec{\Pi}_C \}$. The canonical pair $(\vec{P}, \vec{\Pi}_{P})$ has already been separated and is therefore not included in the summation.

With these definitions, the consistency conditions take the form
\begin{align}
\dot{\Psi}_i
&= M_{ij}\,u_j + v_i \label{Mij+v} = 0 .
\end{align}
where,
\begin{align}
\label{mij_sym}
    M_{ij} u_{j}
=& 
     \int d^{3}y\, 
\Big[
     \frac{\partial}{\partial P_{j} (\mathbf{y})}
     \Big(
   \frac{\partial}{\partial P_{i} (\mathbf{x})}
    \Big \{\frac{1}{k_0}\vec{\Pi}_{C}^2 (\mathbf{x}) -
    \frac{\lambda k_1}{m^2 k_{0}^2} \vec{\Pi}_{C}^2 (\mathbf{x}) - 
    \frac{\lambda k_2}{m^2 k_{0}^2} (\vec{P} (\mathbf{x}) \cdot \vec{\Pi}_{C} (\mathbf{x}) )^2 
    \nonumber \\ &
    -
    V_{\text{eff}} (\mathbf{x})
    \Big \} 
    \Big)
    \Big] u_{j} (\mathbf{y}).
\end{align}

In general, Eq. (\ref{mij_sym}) may possess nonlocal terms in the form of delta functions and their derivatives arising out of the functional derivative with respect to the fields $P_{i}$, which in turn can be integrated out to yield a matrix $M_{ij}$ with components that have local coordinate dependence. Such a matrix can take the following general form:
\begin{align}
\label{mij_gen}
    M_{ij}
=& \alpha_{0}\, \delta_{ij} +
    \alpha_{1} P_{i} P_{j} +
    \alpha_{2} P_{(i} Q_{j)} +
    \alpha_{3} P_{(i} (\Pi_{Q})_{j)}+
    \alpha_{4} P_{(i} \partial_{j)}C_{0} +
    \alpha_{5} P_{(i} C_{j)} +
    \alpha_{6} P_{(i} (\Pi_{C})_{j)}
    \nonumber \\
    &+
    \alpha_{7} Q_{i} Q_{j} +
    \alpha_{8} Q_{(i} (\Pi_{Q})_{j)} 
     +\alpha_{9} Q_{(i} \partial_{j)}C_{0} + 
    \alpha_{10} Q_{(i} C_{j)} +
    \alpha_{11} Q_{(i} (\Pi_{C})_{j)} +
    \alpha_{12} \partial_{i}C_{0} \partial_{j}C_{0} 
    \nonumber \\
    &+
    \alpha_{13} \partial_{(i}C_{0} (\Pi_{Q})_{j)} +
    \alpha_{14} \partial_{(i}C_{0} C_{j)}  +
    \alpha_{15} \partial_{(i}C_{0} (\Pi_{C})_{j)}  +
    \alpha_{16} C_{i} C_{j} +
    \alpha_{17} C_{(i} (\Pi_{Q})_{j)}  
    \nonumber \\
    & +
    \alpha_{18}  C_{(i} (\Pi_{C})_{j)} 
    + \alpha_{19} (\Pi_{Q})_{i} (\Pi_{Q})_{j} +
    \alpha_{20} (\Pi_{Q})_{(i}(\Pi_{C})_{j)} +
    \alpha_{21} (\Pi_{C})_{i}(\Pi_{C})_{j} ,
\end{align}
where $\alpha_{0}, \alpha_1, \dots, \alpha_{21}$ represent generalized coefficients which may constitute scalar functions of the fields.

In principle, the inverse matrix $(M^{-1})_{jk}$ may be obtained using the projection method, wherein one introduces an ansatz for $(M^{-1})_{jk}$ built from the same set of symmetric tensor structures and determines the unknown coefficients by enforcing
\begin{equation}
\label{inveq}
M_{ij}(M^{-1})_{jk} = \delta_{ik} .
\end{equation}
For the class of models under consideration, the structure of $(M^{-1})_{jk}$ will have the form,
\begin{align}
    (M^{-1})_{jk} =&
\beta_{0}\,\delta_{jk}
+ \beta_{1} P_j P_k
+ \beta_{2} P_{(j} Q_{k)}
+ \beta_{3} P_{(j} (\Pi_Q)_{k)}
+ \beta_{4} P_{(j}\partial_{k)} C_0
+ \beta_{5} P_{(j} C_{k)}
+ \beta_{6} P_{(j} (\Pi_C)_{k)} \nonumber \\
&+ \beta_{7} Q_{j} Q_{k}
+ \beta_{8} Q_{(j} (\Pi_Q)_{k)}
+ \beta_{9} Q_{(j}\partial_{k)} C_0
+ \beta_{10} Q_{(j} C_{k)}
+ \beta_{11} Q_{(j} (\Pi_C)_{k)} \nonumber \\
&+ \beta_{12}\,\partial_{j} C_0\,\partial_{k} C_0
+ \beta_{13}\,\partial_{(j} C_0\,(\Pi_Q)_{k)}
+ \beta_{14}\,\partial_{(j} C_0\,C_{k)}
+ \beta_{15}\,\partial_{(j} C_0\,(\Pi_C)_{k)} \nonumber \\
&+ \beta_{16} C_{j} C_{k}
+ \beta_{17} C_{(j} (\Pi_Q)_{k)}
+ \beta_{18} C_{(j} (\Pi_C)_{k)} \nonumber \\
&+ \beta_{19} (\Pi_Q)_{j} (\Pi_Q)_{k}
+ \beta_{20} (\Pi_Q)_{(j} (\Pi_C)_{k)}
+ \beta_{21} (\Pi_C)_{j} (\Pi_C)_{k},
\end{align}
where $\beta_0, \dots, \beta_{21}$ represent the generalized coefficients obtained after inverting the matrix $M_{ij}$. 
For the specific case where $M_{ij}$ is a matrix of local coordinate-dependent components, $M^{-1}$ can also be computed using the standard matrix inversion technique.

\section{$M^{-1}$ for an explicit vacuum structure}
\label{sec4}
In this section, we solve explicitly for the matrix $M_{ij}$ using \textit{Mathematica} \cite{Mathematica}. The resulting expressions are extremely lengthy, typically containing thousands of terms. Because of this complexity, it is not practical to directly compute the inverse of $M_{ij}$ and check whether it becomes singular.
We now consider the explicit potential in Eq. (\ref{eq:pot}) with up to quadratic order derivative terms:
\begin{align}
\label{eq:potquad}
V_{st} = &
(B_{ab}B^{ab})^{2} + 
\frac{1}{m^{2}}\Big(B_{ab}F^{ab}\Big)^{2} + 
\frac{1}{m^{2}}\Big(F_{ab}B^{ab}\Big)^{2} + b^{4} + 
\frac{2}{m} B_{ab} B^{ab} B_{cd} F^{cd} \nonumber \\
& 
+ \frac{2}{m} B_{ab} B^{ab} F_{cd} B^{cd} + 
\frac{2}{m^{2}} B_{ab} B^{ab} F_{cd} F^{cd} 
- 2 b^2 B_{ab} B^{ab} + 
\frac{2}{m^{2}} B_{ab} F^{ab} F_{cd} B^{cd} \nonumber \\
& - 
\frac{2 b^2}{m} B_{ab} F^{ab} - 
\frac{2 b^2}{m} F_{ab} B^{ab} - 
\frac{2 b^2}{m^{2}} F_{ab} F^{ab} .
\end{align}
The Hamiltonian, obtained after carrying out the steps outlined in Sec. \ref{sec3}, is given by,
\begin{align}
    \mathcal{H}_{A} =& \frac{1}{2} \vec{\Pi}_{Q}^{2} +
    \vec{\Pi}_{Q} \cdot (\vec{\nabla} \times \vec{P} )
    +\frac{1}{2}(\vec{\nabla} \cdot \vec{Q})^2
    +\lambda \Big\{ 4 ( \vec{P}^{2} - \vec{Q}^{2})^2 
    + \frac{8}{m^{2}} [\vec{P} \cdot \vec{\Theta} 
+ \vec{P} \cdot \vec{\nabla}C_{0}  \nonumber \\
&
+\vec{Q} \cdot ( \vec{\nabla} \times \vec{C}  )]^{2} 
+ b^{4} - 
\frac{16}{m} [(\vec{P}^{2} - \vec{Q}^{2}) \{\vec{P} \cdot \vec{\Theta} + \vec{P} \cdot \vec{\nabla}C_{0} + \vec{Q} \cdot ( \vec{\nabla} \times \vec{C})\}] \nonumber \\ 
&
+\frac{16}{m^{2}} [(\vec{P}^{2} - \vec{Q}^{2} ) \{ \vec{\Theta}^{2} +(\vec{\nabla}C_{0} )^{2} + 
2 (\vec{\Theta} \cdot \vec{\nabla}C_{0}) - (\vec{\nabla} \times \vec{C} )^{2}\} ] \nonumber \\
& 
+ 4 b^2 ( \vec{P}^{2} - \vec{Q}^{2}) - 
\frac{8 b^2}{m} [\vec{P} \cdot \vec{\Theta} + \vec{Q} \cdot ( \vec{\nabla} \times \vec{C}) + \vec{P} \cdot \vec{\nabla}{C_{0}} ]  \nonumber \\
&
 + \frac{4 b^2}{m^{2}} [\vec{\Theta}^{2} + {2} (\vec{\Theta} \cdot \vec{\nabla}{C_{0}}) + (\vec{\nabla}{C_{0}} )^{2} -  ( \vec{\nabla} \times \vec{C})^{2}  ] \Big\}
     + \vec{u} \cdot  \vec{\Pi}_{P} + 
 w \Pi_{C_{0}} \label{Hamiltonian_H},
\end{align}
where $\vec{\Theta}$ is given by,
\begin{align}
\vec{\Theta}
=\;
\frac{-1}{\lambda \left[
\dfrac{16}{m^{2}} \left(3\vec P^{\,2}-2\vec Q^{\,2}\right)
+\dfrac{8 b^2}{m^{2}} 
\right]}&
 \Bigg\{
\vec\Pi_C
+ \lambda\Bigg[
\frac{16}{m^{2}}
\{\vec P\cdot\vec{\nabla} C_0
+\vec Q\cdot(\vec{\nabla} \times \vec C)\}\vec P
+\frac{8 b^2}{m^{2}}\vec{\nabla} C_0
\nonumber \\
&
-\frac{16}{m}\,(\vec P^{\,2}-\vec Q^{\,2})\,\vec P
+\frac{32}{m^{2}}\,(\vec P^{\,2}-\vec Q^{\,2})\vec{\nabla} C_0
-\frac{8 b^2}{m}\,\vec P
\Bigg]
\Bigg\}. \label{CDot}
\end{align}
As noted earlier in Eq. (\ref{cdot}), the denominator of Eq. (\ref{CDot}) may vanish in a general phase space for specific choices of $\vec{P}$ and $\vec{Q}$, rendering the Hamiltonian (\ref{Hamiltonian_H}) ill-defined. While restricting the physically relevant phase space to the vicinity of the vacuum manifold, defined by the condition $2(\vec{Q}^2-\vec{P}^2)=b^2$---can locally bypass this singularity, consistent canonical quantization relies on a globally well-defined Hamiltonian. The existence of regions in the phase space where the denominator vanishes signifies a fundamental obstruction to globally defining the symplectic structure.

To continue with an explicit calculation, we evaluate $M_{ij}$ on the vacuum manifold defined by choosing the simplest but physically consistent vacuum structure of the antisymmetric tensor field, $B_{ab}\approx b_{ab}$ based on the argument that it is possible to reach a special observer frame— locally in a Lorentz frame in curved Riemann spacetime or globally in Minkowski spacetime— in which the vacuum tensor $b_{ab}$ takes a simple block-diagonal form \cite{Altschul_09},
\begin{equation}
b_{ab}
=
\begin{pmatrix}
0 & -a & 0 & 0 \\
a & 0 & 0 & 0 \\
0 & 0 & 0 & b' \\
0 & 0 & -b' & 0
\end{pmatrix} ,
\label{eq:bmn_block}
\end{equation}
provided that $b_{ab}b^{ab} = -2 (a^2-b'^2)$ is non-zero and $a,b'$ are real constants. Furthermore, the analysis of monopole solutions for antisymmetric tensor fields in Refs.~\cite{Seifert_10-1, Seifert_10-2} shows that a spherically symmetric, nontrivial solution of the equations of motion for $B_{ab}$ that asymptotically approaches the vacuum configuration allows for taking $b'=0$. 

We therefore choose an observer frame in which the background tensor $b_{ab}$ takes a simple canonical form in the vicinity of the vacuum. In this frame, we further restrict ourselves to a representative vacuum configuration in which most of the canonical variables vanish. In particular, all components of $\vec{P}$ and $\vec{Q}$ are set to zero except for a single non-zero component $P_x = -a $. This choice significantly simplifies the algebraic structure of the matrix $M_{ij}$ while preserving the essential physical features of the vacuum configuration.

In practice, this means that in our calculation, we substitute
\begin{equation}
P_x = -a, \qquad P_y = P_z = 0, \qquad \vec{Q} = 0 , \nonumber
\end{equation}
provided that $b_{ab} b^{ab} = -2 \,a^2$ is non-zero and $ a $ is a real constant, and then evaluated the entries of the matrix $M_{ij}$ on this configuration. As a result, we obtain explicit expressions for the individual components of the matrix,
\begin{align}
M_{xx} =
&-\frac{3m}{a} (\Pi_C)_x
+\frac{ m^2}{64 \lambda\,a^4} \ \big\{15 (\Pi_C)_x^2
+28 (\Pi_C)_y^2
+28(\Pi_{C})_z^2 \big \}\nonumber\\[6pt]
&-\frac{3}{a^2}\,\{(\Pi_C)_z \,\partial_z C_{0}
+ (\Pi_C)_y\,\partial_y C_{0} \}
+\frac{32\lambda}{m^2}
( \vec{\nabla } \times \vec{C}
\,)^2 , \nonumber \\
M_{yy} \;=\;
&\frac{m}{a}\,(\Pi_C)_x
-\frac{m^2}{64\lambda\,a^4}\,\big\{ 2 (\Pi_C)_x^2
+ 6 (\Pi_C)_y^2
+  5 (\Pi_C)_z^2 \big \} 
+\frac{1}{a^2}\,(\Pi_C)_y\,\partial_y C_{0}
\nonumber\\[6pt]
&
+\frac{1}{2a^2}\big\{(\Pi_C)_z\,\partial_z C_{0}
-(\Pi_C)_x\,\partial_x C_{0} \big \}
-\frac{4\lambda}{m^2}\Big[(\partial_z C_{0})^2+(\partial_x C_{0})^2\Big]
+\frac{32\lambda}{m^2} \big(\vec{\nabla} \times \vec{C} \big)^2,
\nonumber \\
M_{zz} =\;
&\frac{m}{a}\,(\Pi_C)_x 
-\frac{m^2}{64\,\lambda\,a^4}\big\{2 (\Pi_C)_x^2 
+5 (\Pi_C)_y^2 
+ 6 (\Pi_C)_z^2 \big\} 
+\frac{1}{a^2}\,(\Pi_C)_z \,\partial_z C_{0} 
\nonumber\\[6pt]
&+\frac{1}{2a^2}\big\{(\Pi_C)_y \,\partial_y C_{0} 
+ (\Pi_C)_x \,\partial_x C_{0} \big\} 
-\frac{4\lambda}{m^2}\big\{(\partial_y C_{0} )^2+(\partial_x C_{0} )^2\big\}
+\frac{32\lambda}{m^2} \big(\vec{\nabla} \times \vec{C} \big)^2 , \nonumber
\end{align} 
\begin{align}
M_{xy} =& M_{yx} \;=\;
\frac{m}{a}\,(\Pi_C)_y
+\frac{5 m^2}{64\,\lambda\,a^4}\,(\Pi_C)_x\,(\Pi_C)_y
-\frac{3}{4a^2}\,\big\{(\Pi_C)_x\,\partial_y C_{0}
+ (\Pi_C)_y\,\partial_x C_{0} \big \}
\nonumber\\[6pt]
&+\frac{4\lambda}{m^2}\,
\partial_y C_{0}\,
\partial_x C_{0} \, ,\nonumber \\
M_{xz} =& M_{zx} \;=\;
\frac{m}{a}\,(\Pi_C)_z
+\frac{5 m^2}{64\,\lambda\,a^4}\,(\Pi_C)_x\,(\Pi_C)_z
-\frac{3}{4a^2}\,\big\{(\Pi_C)_x\,\partial_z C_{0}
+ (\Pi_C)_z\,\partial_x C_{0} \big \}
\nonumber\\[6pt]
&+\frac{4\lambda}{m^2}\,
\partial_z C_{0}\,
\partial_x C_{0} \, , \nonumber \\
M_{yz} =& M_{zy} \;=\;
-\frac{m^2}{64\,\lambda\,a^4}\,(\Pi_C)_y \,(\Pi_C)_z
+\frac{1}{4a^2}\big\{(\Pi_C)_y \,\partial_z C_{0} 
+ (\Pi_C)_z \,\partial_y C_{0} \big\}
\nonumber\\[6pt]
&+\frac{4\lambda}{m^2}\,
\partial_z C_{0} \,
\partial_y C_{0} \, ,
\label{Matrix_Components}
\end{align}

The augmented Hamiltonian becomes singular if the inverse matrix $M^{-1}$ is not globally well defined. Following the analysis in Ref.~\cite{Seifert_19}, the divergent terms in $M^{-1}$ arise when the denominator appearing in its components vanishes, which is directly related to the vanishing of the determinant of $M_{ij}$. To examine the invertibility of the constraint matrix defined in Eq.~(\ref{Matrix_Components}), we compute the inverse of $M_{ij}$ for the chosen vacuum configuration (\ref{eq:bmn_block}). The individual components of this inverse matrix are provided in Appendix \ref{appA}.

Since each component of the inverse matrix is proportional to $1/\det(M_{ij})$, the singularity of the Hamiltonian is contingent upon whether the determinant $\det(M_{ij})$ may vanish globally. The determinant evaluated on the vacuum manifold is shown in Appendix \ref{appB}. Upon imposing the vacuum conditions, we find that the determinant exhibits an explicit dependence on the conjugate momenta $\vec{\Pi}_{C}$ and the spatial gradients of the St\"uckelberg fields (i.e., $\nabla_i C_a$). Additionally, all coefficients depend on the Lorentz-violating parameter $a$ through the vacuum structure in Eq. (\ref{eq:bmn_block}). Even if one imposes the unitary gauge condition $C_a = 0$ to recover the original formulation in Eq. (\ref{eq_genL}), the conjugate momenta and spatial gradients do not identically vanish because $C_a$ is a dynamical field. Consequently, the determinant $\det(M_{ij})$ is dynamically contingent. The matrix $M_{ij}$ exhibits dynamical degeneracy over the global phase space; there is no structural guarantee that it remains non-zero. Because a globally valid inverse $(M^{-1})_{ij}$ cannot be established, the fundamental symplectic structure is lost.

\section{Discussions and conclusions}
\label{sec5}
It was shown in Ref. \cite{Seifert_19} that the constraint matrix $M_{ij}$ corresponding to a spontaneous Lorentz-violating model of rank-two antisymmetric tensor field of the form Eq. (\ref{eq_genL}) can become singular on the vacuum manifold regardless of the choice of potential and kinetic terms. Moreover, the nature of this singularity is rather generic, such that it cannot be resolved even in a restricted phase space \cite{Seifert_19}. In this work, we studied the Hamiltonian formulation of a St\"uckelberg-inspired model of a rank-two antisymmetric tensor field wherein the gauge symmetry is restored by introducing a so-called St\"uckelberg vector field $C_a$ while retaining only up to quadratic order derivative terms, and examined the conditions for a singular Hamiltonian on the vacuum manifold to determine if global quantizability can be achieved.

There are two main conclusions of this work: first, the augmented Hamiltonian [Eqs.~(\ref{Hamiltonian_1}) and (\ref{Hamiltonian_H})] is globally ill-defined due to vanishing denominator terms, thus preventing consistent canonical quantization in the full phase space. Second, even in a restricted phase space strictly in an effective field theoretic sense, where the denominator is nonvanishing due to the vacuum manifold conditions, our analysis shows that the determinant of the constraint matrix $M_{ij}$ acquires an explicit dependence on the gradients and conjugate momenta of the St\"uckelberg field components on the vacuum manifold, as shown in Eq. (\ref{mij_sym}). The matrix $M_{ij}$ therefore exhibits dynamical degeneracy, and there is no global structural guarantee that these conjugate momenta $\vec{\Pi}_C$ or field-gradients $\partial_{i}C_{a}$ ($a=0,1,2,3$) remain non-vanishing. 

For completeness, a direct evaluation of $M_{ij}$ and its inverse for an explicit model has been carried out using the Mathematica package, and the notebook with results is hosted on the following link: \url{https://github.com/gomdsaif/Hamiltonian_Constraint_Analysis}. We explicitly evaluated the constraint matrix $M_{ij}$ on a physically motivated vacuum configuration of $B_{ab}=b_{ab}$, where the structure of $b_{ab}$ is given by Eq. (\ref{eq:bmn_block}) \cite{Seifert_10-1, Seifert_10-2}, explicitly confirming this dynamical contingency. 

A few comments related to the physical implications are in order. Models of spontaneous Lorentz violation, such as the antisymmetric tensor model studied here, are effective field theories (EFTs). Their physical validity is strictly restricted to small perturbative fluctuations around the non-zero vacuum expectation value,
$\langle B_{ab} \rangle = b_{ab}$.
If the fields evolve sufficiently far from this background such that the denominator of Eq.~(\ref{CDot}) approaches zero, the EFT approximation has already broken down. Therefore, while global invertibility is not guaranteed, the Hamiltonian can remain locally well defined within the corresponding restricted region of phase space.

The St\"uckelberg mechanism allows the Hamiltonian to remain nonsingular on and strictly near the vacuum manifold in a restricted phase space where the St\"uckelberg field conjugate momenta and their gradients are nonvanishing. This is the regime in which Seifert~\cite{Seifert_19} demonstrated that the original formulation possesses generic pathologies that cannot be resolved even in a restricted phase space. However, this dynamical contingency also offers a constructive perspective for physical applications. This nonsingular behavior is contingent upon the dynamics of the St\"uckelberg field, since the determinant of $M_{ij}$ depends explicitly on its conjugate momenta, and its spatial gradients. Thus, if the dynamical evolution of a particular field configuration remains within a region satisfying $\det(M_{ij})\neq 0$, the inverse constraint matrix exists and a consistent local canonical description can, in principle, be constructed.

Establishing such a consistent theoretical framework is also motivated by the broader interest in theories beyond the standard gravitational framework. Antisymmetric tensor fields naturally arise in the low-energy description of string theory, while their non-zero vacuum expectation values can provide a mechanism for spontaneous Lorentz-symmetry breaking and associated anisotropic effects~\cite{Hernask_16}. Such models are of particular interest in cosmology, where Kalb--Ramond backgrounds have been studied in connection with direction-dependent cosmic expansion and possible anisotropic signatures in the early universe~\cite{Maluf_22, Lessa_19, Atamurotov_22, JuniorEdnaldo_24}. Therefore, for a consistent cosmological application of the present model, it is important to determine whether the evolution of a given cosmological background keeps the St\"uckelberg fields within a nonsingular region of phase space. The present analysis provides a condition for making this assessment and hence offers a useful starting point for investigating consistent anisotropic cosmological solutions in the future.

As noted earlier, an appropriate gauge-fixing procedure would be required a priori to deal with the currently undetermined Lagrange multiplier $w$. However, the failure of structural invertibility in $M_{ij}$ occurs prior to this gauge-fixing step and would not impact the present conclusions. This work also provides critical context to the recent work by Potting \cite{Potting_23}, where a Lorentz-violating 'hybrid' model of an antisymmetric tensor field coupled to a vector field was constructed and found to be bounded from below. Our canonical analysis indicates that while such hybrid constructions might exhibit classical stability, introducing a vector field via the St\"uckelberg mechanism does not structurally guarantee the global non-singularity required for standard canonical quantization.


\newpage
\appendix

\section{Elements of the inverse matrix}
\label{appA}
Here we have shown the elements of the inverse of the matrix $M_{ij}$

\begin{align}
(M^{-1})_{xx} &= \frac{1}{\det(M_{ij})} \Bigg( -\bigg(
-\frac{m^2}{64 \lambda a^4} (\Pi_C)_y (\Pi_C)_z
+ \frac{1}{4 a^2} (\Pi_C)_y\,\partial_z C_0
+ \frac{1}{4 a^2} (\Pi_C)_z\,\partial_y C_0  \nonumber \\
& 
+ \frac{4 \lambda}{m^2} \partial_z C_0  \partial_y C_0
\bigg)^2 
+
\Bigg(
\frac{m}{a} (\Pi_C)_x 
- \frac{m^2}{32 \lambda a^4} (\Pi_C)_x^2
- \frac{3 m^2}{32 \lambda a^4} (\Pi_C)_y^2
- \frac{5 m^2}{64 \lambda a^4} (\Pi_C)_z^2 \nonumber \\
& 
+ \frac{1}{2 a^2} (\Pi_C)_z\,\partial_z C_0 
- \frac{4 \lambda}{m^2} (\partial_z C_0)^2 
+ \frac{32 \lambda}{m^2} (\partial_z C_x)^2
+ \frac{32 \lambda}{m^2} (\partial_z C_y)^2 
+ \frac{1}{a^2} (\Pi_C)_y\,\partial_y C_0 \nonumber \\
& 
+ \frac{32 \lambda}{m^2} (\partial_y C_x)^2 
- \frac{64 \lambda}{m^2} (\partial_z C_y)(\partial_y C_z)
+ \frac{32 \lambda}{m^2} (\partial_y C_z)^2 
- \frac{1}{2 a^2} (\Pi_C)_x\,\partial_x C_0  
- \frac{4 \lambda}{m^2} (\partial_x C_0)^2 \nonumber \\
& 
- \frac{64 \lambda}{m^2} (\partial_y C_x)(\partial_x C_y)
+ \frac{32 \lambda}{m^2} (\partial_x C_y)^2
- \frac{64 \lambda}{m^2} (\partial_z C_x)(\partial_x C_z)  
+ \frac{32 \lambda}{m^2} (\partial_x C_z)^2  
\Bigg)  
\Bigg(
\frac{m}{a} (\Pi_C)_x \nonumber \\
& 
- \frac{m^2}{32 \lambda a^4} (\Pi_C)_x^2
- \frac{5 m^2}{64 \lambda a^4} (\Pi_C)_y^2
- \frac{3 m^2}{32 \lambda a^4} (\Pi_C)_z^2
+ \frac{1}{a^2} (\Pi_C)_z\,\partial_z C_0 
+ \frac{32 \lambda}{m^2} (\partial_z C_x)^2 \nonumber \\
& 
+ \frac{32 \lambda}{m^2} (\partial_z C_y)^2
+ \frac{1}{2 a^2} (\Pi_C)_y\,\partial_y C_0
- \frac{4 \lambda}{m^2} (\partial_y C_0)^2
+ \frac{32 \lambda}{m^2} (\partial_y C_x)^2 
- \frac{64 \lambda}{m^2} (\partial_z C_y)(\partial_y C_z) \nonumber \\
& 
+ \frac{32 \lambda}{m^2} (\partial_y C_z)^2
- \frac{1}{2 a^2} (\Pi_C)_x\,\partial_x C_0
- \frac{4 \lambda}{m^2} (\partial_x C_0)^2
- \frac{64 \lambda}{m^2} (\partial_y C_x)(\partial_x C_y)  
+ \frac{32 \lambda}{m^2} (\partial_x C_y)^2 \nonumber \\
& 
- \frac{64 \lambda}{m^2} (\partial_z C_x)(\partial_x C_z)
+ \frac{32 \lambda}{m^2} (\partial_x C_z)^2
\Bigg) \Bigg) 
\nonumber,
\end{align}

\begin{align}
   (M^{-1})_{yy} &=\frac{1}{\det(M_{ij})}
   \Bigg( 
-\bigg(
\frac{m}{a} (\Pi_C)_z
+ \frac{5 m^2}{64 \lambda a^4} (\Pi_C)_x (\Pi_C)_z
- \frac{3}{4 a^2} (\Pi_C)_x \partial_z C_0
- \frac{3}{4 a^2} (\Pi_C)_z \partial_x C_0 \nonumber \\
& 
+ \frac{4 \lambda}{m^2} \partial_z C_0 \partial_x C_0
\bigg)^2
+
\bigg(
-\frac{3 m}{a} (\Pi_C)_x
+ \frac{15 m^2}{64 \lambda a^4} (\Pi_C)_x^2
+ \frac{7 m^2}{16 \lambda a^4} (\Pi_C)_y^2
+ \frac{7 m^2}{16 \lambda a^4} (\Pi_C)_z^2 \nonumber \\
&
- \frac{3}{a^2} (\Pi_C)_z\,\partial_z C_0 
+ \frac{32 \lambda}{m^2} (\partial_z C_x)^2  
+ \frac{32 \lambda}{m^2} (\partial_z C_y)^2
- \frac{3}{a^2} (\Pi_C)_y\,\partial_y C_0
+ \frac{32 \lambda}{m^2} (\partial_y C_x)^2 \nonumber \\
&
- \frac{64 \lambda}{m^2} \partial_z C_y \partial_y C_z
+ \frac{32 \lambda}{m^2} (\partial_y C_z)^2
- \frac{64 \lambda}{m^2} \partial_y C_x \partial_x C_y
+ \frac{32 \lambda}{m^2} (\partial_x C_y)^2
- \frac{64 \lambda}{m^2} \partial_z C_x \partial_x C_z \nonumber \\
&
+ \frac{32 \lambda}{m^2} (\partial_x C_z)^2
\bigg)
\bigg(
\frac{m}{a} (\Pi_C)_x
- \frac{m^2}{32 \lambda a^4} (\Pi_C)_x^2
- \frac{5 m^2}{64 \lambda a^4} (\Pi_C)_y^2
- \frac{3 m^2}{32 \lambda a^4} (\Pi_C)_z^2 \nonumber \\
&
+ \frac{1}{a^2} (\Pi_C)_z\,\partial_z C_0
+ \frac{32 \lambda}{m^2} (\partial_z C_x)^2 
+ \frac{32 \lambda}{m^2} (\partial_z C_y)^2 
+ \frac{1}{2 a^2} (\Pi_C)_y\,\partial_y C_0
- \frac{4 \lambda}{m^2} (\partial_y C_0)^2 \nonumber \\
&
+ \frac{32 \lambda}{m^2} (\partial_y C_x)^2
- \frac{64 \lambda}{m^2} \partial_z C_y \partial_y C_z
+ \frac{32 \lambda}{m^2} (\partial_y C_z)^2  
- \frac{1}{2 a^2} (\Pi_C)_x\,\partial_x C_0
- \frac{4 \lambda}{m^2} (\partial_x C_0)^2 \nonumber \\
&
- \frac{64 \lambda}{m^2} \partial_y C_x \partial_x C_y
+ \frac{32 \lambda}{m^2} (\partial_x C_y)^2
- \frac{64 \lambda}{m^2} \partial_z C_x \partial_x C_z
+ \frac{32 \lambda}{m^2} (\partial_x C_z)^2
\bigg)
\Bigg), \nonumber
\end{align}

\begin{align}
     (M^{-1})_{zz} = & \frac{1}{\det(M_{ij})}
   \Bigg(
-\bigg(
\frac{m}{a} (\Pi_C)_y
+ \frac{5 m^2}{64 \lambda a^4} (\Pi_C)_x (\Pi_C)_y
- \frac{3}{4 a^2} (\Pi_C)_x\,\partial_y C_0
- \frac{3}{4 a^2} (\Pi_C)_y\,\partial_x C_0 \nonumber \\
&
+ \frac{4 \lambda}{m^2} (\partial_y C_0)(\partial_x C_0)
\bigg)^2 
+
\bigg(
-\frac{3 m}{a} (\Pi_C)_x
+ \frac{15 m^2}{64 \lambda a^4} (\Pi_C)_x^2
+ \frac{7 m^2}{16 \lambda a^4} (\Pi_C)_y^2 \nonumber \\
&
+ \frac{7 m^2}{16 \lambda a^4} (\Pi_C)_z^2
- \frac{3}{a^2} (\Pi_C)_z\,\partial_z C_0
+ \frac{32 \lambda}{m^2} (\partial_z C_x)^2 
+ \frac{32 \lambda}{m^2} (\partial_z C_y)^2
- \frac{3}{a^2} (\Pi_C)_y\,\partial_y C_0 \nonumber \\
&
+ \frac{32 \lambda}{m^2} (\partial_y C_x)^2
- \frac{64 \lambda}{m^2} (\partial_z C_y)(\partial_y C_z)
+ \frac{32 \lambda}{m^2} (\partial_y C_z)^2 
- \frac{64 \lambda}{m^2} (\partial_y C_x)(\partial_x C_y) \nonumber \\
&
+ \frac{32 \lambda}{m^2} (\partial_x C_y)^2 
- \frac{64 \lambda}{m^2} (\partial_z C_x)(\partial_x C_z)
+ \frac{32 \lambda}{m^2} (\partial_x C_z)^2
\bigg)
\bigg(
\frac{m}{a} (\Pi_C)_x 
- \frac{m^2}{32 \lambda a^4} (\Pi_C)_x^2 \nonumber \\
&
- \frac{3 m^2}{32 \lambda a^4} (\Pi_C)_y^2
- \frac{5 m^2}{64 \lambda a^4} (\Pi_C)_z^2
+ \frac{1}{2 a^2} (\Pi_C)_z\,\partial_z C_0
- \frac{4 \lambda}{m^2} (\partial_z C_0)^2
+ \frac{32 \lambda}{m^2} (\partial_z C_x)^2  \nonumber \\
&
+ \frac{32 \lambda}{m^2} (\partial_z C_y)^2 
+ \frac{1}{a^2} (\Pi_C)_y\,\partial_y C_0
+ \frac{32 \lambda}{m^2} (\partial_y C_x)^2
- \frac{64 \lambda}{m^2} (\partial_z C_y)(\partial_y C_z) \nonumber \\
&
+ \frac{32 \lambda}{m^2} (\partial_y C_z)^2 
- \frac{1}{2 a^2} (\Pi_C)_x\,\partial_x C_0
- \frac{4 \lambda}{m^2} (\partial_x C_0)^2
- \frac{64 \lambda}{m^2} (\partial_y C_x)(\partial_x C_y)
+ \frac{32 \lambda}{m^2} (\partial_x C_y)^2 \nonumber \\
&
- \frac{64 \lambda}{m^2} (\partial_z C_x)(\partial_x C_z) 
+ \frac{32 \lambda}{m^2} (\partial_x C_z)^2
\bigg)
\Bigg), \nonumber
\end{align}

\begin{align}
(M^{-1})_{xy} & =(M^{-1})_{yx}=  \frac{1}{\det(M_{ij})} \Bigg( 
\bigg(
-\frac{m^2}{64 \lambda a^4} (\Pi_C)_y (\Pi_C)_z
+ \frac{1}{4 a^2} (\Pi_C)_y\,\partial_z C_0
+ \frac{1}{4 a^2} (\Pi_C)_z\,\partial_y C_0 \nonumber \\
&
+ \frac{4 \lambda}{m^2} \partial_z C_0 \partial_y C_0
\bigg) 
\bigg(
\frac{m}{a} (\Pi_C)_z 
+ \frac{5 m^2}{64 \lambda a^4} (\Pi_C)_x (\Pi_C)_z 
- \frac{3}{4 a^2} (\Pi_C)_x\,\partial_z C_0
- \frac{3}{4 a^2} (\Pi_C)_z\,\partial_x C_0\nonumber \\
&
+ \frac{4 \lambda}{m^2} \partial_z C_0 \partial_x C_0
\bigg)
-
\bigg(
\frac{m}{a} (\Pi_C)_y 
+ \frac{5 m^2}{64 \lambda a^4} (\Pi_C)_x (\Pi_C)_y 
- \frac{3}{4 a^2} (\Pi_C)_x\,\partial_y C_0 \nonumber \\
&
- \frac{3}{4 a^2} (\Pi_C)_y\,\partial_x C_0
+ \frac{4 \lambda}{m^2} \partial_y C_0 \partial_x C_0
\bigg)
\bigg(
\frac{m}{a} (\Pi_C)_x
- \frac{m^2}{32 \lambda a^4} (\Pi_C)_x^2 
- \frac{5 m^2}{64 \lambda a^4} (\Pi_C)_y^2 \nonumber \\
&
- \frac{3 m^2}{32 \lambda a^4} (\Pi_C)_z^2
+ \frac{1}{a^2} (\Pi_C)_z\,\partial_z C_0
+ \frac{32 \lambda}{m^2} (\partial_z C_x)^2
+ \frac{32 \lambda}{m^2} (\partial_z C_y)^2
+ \frac{1}{2 a^2} (\Pi_C)_y\,\partial_y C_0 \nonumber \\
&
- \frac{4 \lambda}{m^2} (\partial_y C_0)^2 
+ \frac{32 \lambda}{m^2} (\partial_y C_x)^2 
- \frac{64 \lambda}{m^2} \partial_z C_y \partial_y C_z
+ \frac{32 \lambda}{m^2} (\partial_y C_z)^2
- \frac{1}{2 a^2} (\Pi_C)_x\,\partial_x C_0 \nonumber \\
&
- \frac{4 \lambda}{m^2} (\partial_x C_0)^2
- \frac{64 \lambda}{m^2} \partial_y C_x \partial_x C_y 
+ \frac{32 \lambda}{m^2} (\partial_x C_y)^2
- \frac{64 \lambda}{m^2} \partial_z C_x \partial_x C_z
+ \frac{32 \lambda}{m^2} (\partial_x C_z)^2
\bigg)
\Bigg), \nonumber
\end{align}
\begin{align}
    (M^{-1})_{xz} &= (M^{-1})_{zx}=  \frac{1}{\det(M_{ij})} 
    \Bigg(  
    \bigg(
-\frac{m^2}{64 \lambda a^4} (\Pi_C)_y (\Pi_C)_z
+ \frac{1}{4 a^2} (\Pi_C)_y\,\partial_z C_0
+ \frac{1}{4 a^2} (\Pi_C)_z\,\partial_y C_0 \nonumber \\
&
+ \frac{4 \lambda}{m^2} \partial_z C_0 \partial_y C_0
\bigg) 
\bigg(
\frac{m}{a} (\Pi_C)_y 
+ \frac{5 m^2}{64 \lambda a^4} (\Pi_C)_x (\Pi_C)_y
- \frac{3}{4 a^2} (\Pi_C)_x\,\partial_y C_0 
- \frac{3}{4 a^2} (\Pi_C)_y\,\partial_x C_0 \nonumber \\
&
+ \frac{4 \lambda}{m^2} \partial_y C_0 \partial_x C_0
\bigg)
-
\bigg(
\frac{m}{a} (\Pi_C)_z 
+ \frac{5 m^2}{64 \lambda a^4} (\Pi_C)_x (\Pi_C)_z
- \frac{3}{4 a^2} (\Pi_C)_x\,\partial_z C_0 \nonumber \\
&
- \frac{3}{4 a^2} (\Pi_C)_z\,\partial_x C_0
+ \frac{4 \lambda}{m^2} \partial_z C_0 \partial_x C_0
\bigg) 
\bigg(
\frac{m}{a} (\Pi_C)_x
- \frac{m^2}{32 \lambda a^4} (\Pi_C)_x^2 
- \frac{3 m^2}{32 \lambda a^4} (\Pi_C)_y^2 \nonumber \\
&
- \frac{5 m^2}{64 \lambda a^4} (\Pi_C)_z^2 
+ \frac{1}{2 a^2} (\Pi_C)_z\,\partial_z C_0
- \frac{4 \lambda}{m^2} (\partial_z C_0)^2
+ \frac{32 \lambda}{m^2} (\partial_z C_x)^2
+ \frac{32 \lambda}{m^2} (\partial_z C_y)^2 \nonumber \\
&
+ \frac{1}{a^2} (\Pi_C)_y\,\partial_y C_0  
+ \frac{32 \lambda}{m^2} (\partial_y C_x)^2 
- \frac{64 \lambda}{m^2} \partial_z C_y \partial_y C_z
+ \frac{32 \lambda}{m^2} (\partial_y C_z)^2
- \frac{1}{2 a^2} (\Pi_C)_x\,\partial_x C_0 \nonumber \\
&
- \frac{4 \lambda}{m^2} (\partial_x C_0)^2 
- \frac{64 \lambda}{m^2} \partial_y C_x \partial_x C_y
+ \frac{32 \lambda}{m^2} (\partial_x C_y)^2
- \frac{64 \lambda}{m^2} \partial_z C_x \partial_x C_z
+ \frac{32 \lambda}{m^2} (\partial_x C_z)^2
\bigg)
    \Bigg) , \nonumber
\end{align}
\begin{align}
(M^{-1})_{yz} & = (M^{-1})_{zy} = \frac{1}{\det(M_{ij})}
   \Bigg(
\bigg(
\frac{m}{a} (\Pi_C)_z
+ \frac{5 m^2}{64 \lambda a^4} (\Pi_C)_x (\Pi_C)_z
- \frac{3}{4 a^2} (\Pi_C)_x\,\partial_z C_0 \nonumber \\
&
- \frac{3}{4 a^2} (\Pi_C)_z\,\partial_x C_0 
+ \frac{4 \lambda}{m^2} \partial_z C_0 \partial_x C_0
\bigg) 
\bigg(
\frac{m}{a} (\Pi_C)_y 
+ \frac{5 m^2}{64 \lambda a^4} (\Pi_C)_x (\Pi_C)_y \nonumber \\
&
- \frac{3}{4 a^2} (\Pi_C)_x\,\partial_y C_0 
- \frac{3}{4 a^2} (\Pi_C)_y\,\partial_x C_0
+ \frac{4 \lambda}{m^2} \partial_y C_0 \partial_x C_0
\bigg)
-
\bigg(
-\frac{m^2}{64 \lambda a^4} (\Pi_C)_y (\Pi_C)_z  \nonumber \\
&
+ \frac{1}{4 a^2} (\Pi_C)_y\,\partial_z C_0 
+ \frac{1}{4 a^2} (\Pi_C)_z\,\partial_y C_0
+ \frac{4 \lambda}{m^2} \partial_z C_0 \partial_y C_0
\bigg)
\bigg(
-\frac{3 m}{a} (\Pi_C)_x \nonumber \\
&
+ \frac{15 m^2}{64 \lambda a^4} (\Pi_C)_x^2
+ \frac{7 m^2}{16 \lambda a^4} (\Pi_C)_y^2 
+ \frac{7 m^2}{16 \lambda a^4} (\Pi_C)_z^2
- \frac{3}{a^2} (\Pi_C)_z\,\partial_z C_0 
+ \frac{32 \lambda}{m^2} (\partial_z C_x)^2 \nonumber \\
&
+ \frac{32 \lambda}{m^2} (\partial_z C_y)^2 
- \frac{3}{a^2} (\Pi_C)_y\,\partial_y C_0 
+ \frac{32 \lambda}{m^2} (\partial_y C_x)^2
- \frac{64 \lambda}{m^2} \partial_z C_y \partial_y C_z
+ \frac{32 \lambda}{m^2} (\partial_y C_z)^2 \nonumber \\
&
- \frac{64 \lambda}{m^2} \partial_y C_x \partial_x C_y
+ \frac{32 \lambda}{m^2} (\partial_x C_y)^2 
- \frac{64 \lambda}{m^2} \partial_z C_x \partial_x C_z
+ \frac{32 \lambda}{m^2} (\partial_x C_z)^2
\bigg)
\Bigg) .
\end{align}

\section{Determinant of the constraint matrix}
\label{appB}
The determinant of the matrix $M_{ij}$ is given as,
\begin{align}
\det(M_{ij}) &=\; 
\bigg( -
\bigg(
\bigg(
-\frac{m^2}{64 \lambda a^4} (\Pi_C)_y (\Pi_C)_z
+ \frac{1}{4 a^2} (\Pi_C)_y\,\partial_z C_0
+ \frac{1}{4 a^2} (\Pi_C)_z\,\partial_y C_0 \nonumber \\
&
+ \frac{4 \lambda}{m^2} \partial_z C_0 \partial_y C_0
\bigg) 
\bigg(- \bigg(
\bigg(
\frac{m}{a} (\Pi_C)_z 
+ \frac{5 m^2}{64 \lambda a^4} (\Pi_C)_x (\Pi_C)_z
- \frac{3}{4 a^2} (\Pi_C)_x\,\partial_z C_0 \nonumber \\
&
- \frac{3}{4 a^2} (\Pi_C)_z\,\partial_x C_0
+ \frac{4 \lambda}{m^2} \partial_z C_0 \partial_x C_0 
\bigg)
\bigg(
\frac{m}{a} (\Pi_C)_y
+ \frac{5 m^2}{64 \lambda a^4} (\Pi_C)_x (\Pi_C)_y  \nonumber \\
&
- \frac{3}{4 a^2} (\Pi_C)_x\,\partial_y C_0 
- \frac{3}{4 a^2} (\Pi_C)_y\,\partial_x C_0 
+ \frac{4 \lambda}{m^2}  \partial_y C_0 \partial_x C_0 
\bigg) \bigg)
+
\bigg(
-\frac{m^2}{64 \lambda a^4} (\Pi_C)_y (\Pi_C)_z \nonumber \\
&
+ \frac{1}{4 a^2} (\Pi_C)_y\,\partial_z C_0  
+ \frac{1}{4 a^2} (\Pi_C)_z\,\partial_y C_0 
+ \frac{4 \lambda}{m^2} \partial_z C_0 \partial_y C_0 
\bigg)
\bigg(
-\frac{3m}{a} (\Pi_C)_x
+ \frac{15 m^2}{64 \lambda a^4} (\Pi_C)_x^2 \nonumber \\
&
+ \frac{7 m^2}{16 \lambda a^4} (\Pi_C)_y^2 
+ \frac{7 m^2}{16 \lambda a^4} (\Pi_C)_z^2 
- \frac{3}{a^2} (\Pi_C)_z\,\partial_z C_0
- \frac{3}{a^2} (\Pi_C)_y\,\partial_y C_0
+ \frac{32 \lambda}{m^2} (\partial_z C_x)^2 \nonumber \\
&
+ \frac{32 \lambda}{m^2} (\partial_z C_y)^2 
- \frac{64 \lambda}{m^2} \partial_z C_y \partial_y C_z
+ \frac{32 \lambda}{m^2} (\partial_y C_x)^2
+ \frac{32 \lambda}{m^2} (\partial_y C_z)^2
- \frac{64 \lambda}{m^2} \partial_y C_x \partial_x C_y  \nonumber \\
&
- \frac{64 \lambda}{m^2} \partial_z C_x \partial_x C_z 
+ \frac{32 \lambda}{m^2} (\partial_x C_y)^2 
+ \frac{32 \lambda}{m^2} (\partial_x C_z)^2
\bigg) \bigg) \bigg) 
+
\bigg(
\frac{m}{a} (\Pi_C)_z
+ \frac{5 m^2}{64 \lambda a^4} (\Pi_C)_x (\Pi_C)_z\nonumber \\
&
- \frac{3}{4 a^2} (\Pi_C)_x\,\partial_z C_0 
- \frac{3}{4 a^2} (\Pi_C)_z\,\partial_x C_0 
+ \frac{4 \lambda}{m^2} \partial_z C_0 \partial_x C_0
\bigg) \bigg( \bigg(
-\frac{m^2}{64 \lambda a^4} (\Pi_C)_y (\Pi_C)_z \nonumber \\
&
+ \frac{1}{4 a^2} (\Pi_C)_y\,\partial_z C_0 
+ \frac{1}{4 a^2} (\Pi_C)_z\,\partial_y C_0
+ \frac{4 \lambda}{m^2} \partial_z C_0 \partial_y C_0
\bigg)
\bigg(
\frac{m}{a} (\Pi_C)_y
+ \frac{5 m^2}{64 \lambda a^4} (\Pi_C)_x (\Pi_C)_y \nonumber \\
&
- \frac{3}{4 a^2} (\Pi_C)_x\,\partial_y C_0 
- \frac{3}{4 a^2} (\Pi_C)_y\,\partial_x C_0
+ \frac{4 \lambda}{m^2} \partial_y C_0 \partial_x C_0
\bigg) 
-
\bigg(
\frac{m}{a} (\Pi_C)_z  
\nonumber \\
&
+ \frac{5 m^2}{64 \lambda a^4} (\Pi_C)_x (\Pi_C)_z
- \frac{3}{4 a^2} (\Pi_C)_x\,\partial_z C_0
- \frac{3}{4 a^2} (\Pi_C)_z\,\partial_x C_0
+ \frac{4 \lambda}{m^2} \partial_z C_0 \partial_x C_0
\bigg) 
\bigg(
\frac{m}{a} (\Pi_C)_x \nonumber \\
&
- \frac{m^2}{32 \lambda a^4} (\Pi_C)_x^2 
- \frac{3 m^2}{32 \lambda a^4} (\Pi_C)_y^2
- \frac{5 m^2}{64 \lambda a^4} (\Pi_C)_z^2
+ \frac{1}{2 a^2} (\Pi_C)_z\,\partial_z C_0
+ \frac{1}{a^2} (\Pi_C)_y\,\partial_y C_0 \nonumber \\
&
- \frac{1}{2 a^2} (\Pi_C)_x\,\partial_x C_0 
- \frac{4 \lambda}{m^2} (\partial_z C_0)^2 
- \frac{4 \lambda}{m^2} (\partial_x C_0)^2 
+ \frac{32 \lambda}{m^2} (\partial_z C_x)^2
+ \frac{32 \lambda}{m^2} (\partial_z C_y)^2
+ \frac{32 \lambda}{m^2} (\partial_y C_x)^2 \nonumber \\
&
+ \frac{32 \lambda}{m^2} (\partial_y C_z)^2 
+ \frac{32 \lambda}{m^2} (\partial_x C_y)^2
+ \frac{32 \lambda}{m^2} (\partial_x C_z)^2
- \frac{64 \lambda}{m^2} \partial_z C_y \partial_y C_z
- \frac{64 \lambda}{m^2} \partial_y C_x \partial_x C_y \nonumber \\
&
- \frac{64 \lambda}{m^2} \partial_z C_x \partial_x C_z
\bigg)
\bigg) 
+
\bigg(
\frac{m}{a} (\Pi_C)_x
- \frac{m^2}{32 \lambda a^4} (\Pi_C)_x^2
- \frac{5 m^2}{64 \lambda a^4} (\Pi_C)_y^2
- \frac{3 m^2}{32 \lambda a^4} (\Pi_C)_z^2 \nonumber \\
&
+ \frac{1}{a^2} (\Pi_C)_z\,\partial_z C_0 
+ \frac{1}{2 a^2} (\Pi_C)_y\,\partial_y C_0 
- \frac{1}{2 a^2} (\Pi_C)_x\,\partial_x C_0
- \frac{4 \lambda}{m^2} (\partial_y C_0)^2
- \frac{4 \lambda}{m^2} (\partial_x C_0)^2\nonumber \\
&
+ \frac{32 \lambda}{m^2} (\partial_z C_x)^2 
+ \frac{32 \lambda}{m^2} (\partial_z C_y)^2
+ \frac{32 \lambda}{m^2} (\partial_y C_x)^2 
+ \frac{32 \lambda}{m^2} (\partial_y C_z)^2
+ \frac{32 \lambda}{m^2} (\partial_x C_y)^2
+ \frac{32 \lambda}{m^2} (\partial_x C_z)^2 \nonumber
\end{align}
\begin{align}
&
- \frac{64 \lambda}{m^2} \partial_z C_y \partial_y C_z
- \frac{64 \lambda}{m^2} \partial_y C_x \partial_x C_y
- \frac{64 \lambda}{m^2} \partial_z C_x \partial_x C_z 
\bigg)
\bigg( - 
\bigg(
\frac{m}{a} (\Pi_C)_y
+ \frac{5 m^2}{64 \lambda a^4} (\Pi_C)_x (\Pi_C)_y\nonumber \\
&
- \frac{3}{4 a^2} (\Pi_C)_x\,\partial_y C_0
- \frac{3}{4 a^2} (\Pi_C)_y\,\partial_x C_0
+ \frac{4 \lambda}{m^2} \partial_y C_0 \partial_x C_0
\bigg)^2
+ 
\bigg(
-\frac{3 m}{a} (\Pi_C)_x
+ \frac{15 m^2}{64 \lambda a^4} (\Pi_C)_x^2 \nonumber \\
&
+ \frac{7 m^2}{16 \lambda a^4} (\Pi_C)_y^2
+ \frac{7 m^2}{16 \lambda a^4} (\Pi_C)_z^2
- \frac{3}{a^2} (\Pi_C)_z\,\partial_z C_0 
- \frac{3}{a^2} (\Pi_C)_y\,\partial_y C_0
+ \frac{32 \lambda}{m^2} (\partial_z C_x)^2 \nonumber \\
&
+ \frac{32 \lambda}{m^2} (\partial_z C_y)^2
+ \frac{32 \lambda}{m^2} (\partial_y C_x)^2
+ \frac{32 \lambda}{m^2} (\partial_y C_z)^2 
+ \frac{32 \lambda}{m^2} (\partial_x C_y)^2
+ \frac{32 \lambda}{m^2} (\partial_x C_z)^2 \nonumber \\
&
- \frac{64 \lambda}{m^2} \partial_z C_y \partial_y C_z
- \frac{64 \lambda}{m^2} \partial_y C_x \partial_x C_y 
- \frac{64 \lambda}{m^2} \partial_z C_x \partial_x C_z
\bigg) 
\bigg(
\frac{m}{a} (\Pi_C)_x
- \frac{m^2}{32 \lambda a^4} (\Pi_C)_x^2 \nonumber \\
&
- \frac{3 m^2}{32 \lambda a^4} (\Pi_C)_y^2
- \frac{5 m^2}{64 \lambda a^4} (\Pi_C)_z^2
+ \frac{1}{2a^2} (\Pi_C)_z\,\partial_z C_0 
+ \frac{1}{a^2} (\Pi_C)_y\,\partial_y C_0
- \frac{1}{2a^2} (\Pi_C)_x\,\partial_x C_0 \nonumber \\
&
- \frac{4\lambda}{m^2} (\partial_z C_0)^2
- \frac{4\lambda}{m^2} (\partial_y C_0)^2
- \frac{4\lambda}{m^2} (\partial_x C_0)^2
+ \frac{32\lambda}{m^2} (\partial_z C_x)^2 
+ \frac{32\lambda}{m^2} (\partial_z C_y)^2\nonumber \\
&
+ \frac{32\lambda}{m^2} (\partial_y C_x)^2
+ \frac{32\lambda}{m^2} (\partial_y C_z)^2
+ \frac{32\lambda}{m^2} (\partial_x C_y)^2
+ \frac{32\lambda}{m^2} (\partial_x C_z)^2
- \frac{64\lambda}{m^2} (\partial_z C_y)(\partial_y C_z) \nonumber \\
&
- \frac{64\lambda}{m^2} (\partial_y C_x)(\partial_x C_y)
- \frac{64\lambda}{m^2} (\partial_z C_x)(\partial_x C_z)
\bigg) \bigg )
\bigg).
\end{align}

\bibliographystyle{unsrt}
\bibliography{Ref}

\end{document}